\documentclass[epj]{svjour}

\pdfoutput=1

\usepackage{graphics}
\usepackage{amssymb}
\usepackage[usenames]{color}

\newcommand{\be}{\begin{equation}}
\newcommand{\ee}{\end{equation}}
\newcommand{\bea}{\begin{eqnarray}}
\newcommand{\eea}{\end{eqnarray}}
\newcommand{\bfig}{\begin{figure}}
\newcommand{\efig}{\end{figure}}
\newcommand{\bc}{\begin{center}}
\newcommand{\ec}{\end{center}}
\newcommand{\btab}{\begin{tabular}}
\newcommand{\etab}{\end{tabular}}

\newcommand{\dr}{\partial}

\let\oldepsilon\epsilon
\let\epsilon\varepsilon
\let\varepsilon\oldepsilon

\newcommand{\dx}{\partial_x}

\newcommand{\dz}{\partial_z}
\newcommand{\dt}{\partial_t}

\begin{document}
\title{On the different regimes of subaqueous transport}
\author{B. Andreotti, P. Claudin}
\institute{Laboratoire de Physique et M\'ecanique des Milieux H\'et\'erog\`enes\\
PMMH UMR 7636 CNRS-ESPCI-P6-P7,\\
10 rue Vauquelin, 75231 Paris Cedex 05, France.}
\date{}
\abstract{
We review different aspects of subaqueous sediment transport. We discuss the static threshold and its dependency with longitudinal and transverse slopes, as well as cohesion. We describe the different regimes of transport: erosion and momentum limited bed load, and suspended load. In all these cases, we derive the expressions of the saturation flux $q_{\rm sat}$ and the saturation length $L_{\rm sat}$ and discuss their dependencies.
\PACS{
      {PACS-key}{}   \and
      {PACS-key}{}
     } 
} 
\maketitle
%

\section{Introduction: the saturation length paradigm}
Whenever a sedimentary bed is sheared by a water flow of sufficient strength, the particles at the surface are entrained into motion. The loss of static equilibrium of a grain at the surface of the bed occurs for a value of the driving force $F$ proportional to the submerged weight of the grain: $F \propto ( \rho_s - \rho_f )g d^3$, where $d$ is the grain diameter, $\rho_s$ is the mass density of the sand grain, and $\rho_f$ is that of the fluid. As $F$ is proportional to the shear force $\tau d^2$ exerted by the fluid ($\tau \equiv \rho_f u_*^2$ is the basal shear stress), the non-dimensional parameter controlling the onset of motion is the Shields number $\Theta$, which characterises the ratio of the driving shear stress to the normal stress:
\begin{equation}
\Theta = \frac{\rho_f u_*^2}{(\rho_s-\rho_f)gd}.
\label{Shields}
\end{equation}

When the bed is composed by non-cohesive large grains, phenomenological transport laws relate the sediment flux $q$ --~i.e. the volume of grains crossing a unit transverse length per unit time~-- to the basal shear stress $\tau$. When the bed is composed of small cohesive or cemented particles, one rather uses erosion laws, which relate the erosion rate $\varphi$ --~i.e. the volume of grains leaving a unit bed surface per unit time -- to $\tau$. These two different cases can be described into the same general framework, introducing the concept of saturation length. A relation between the flux and the shear stress is possible if an equilibrium state is reached, e.g. in a steady and homogeneous situation. Hereafter we call $q_{\rm sat}$ the equilibrium or `saturated' sediment flux. $q_{\rm sat}$ is then a function of $\tau$. Any change of the basal shear stress, for instance due to a non flat bed profile, leads to an actual flux which differs from $q_{\rm sat}$, and which generically relaxes towards equilibrium over some distance that we call $L_{\rm sat}$. Then, a situation for which the actual flux is given by a transport law corresponds to the limit $L_{\rm sat} \to 0$, whereas a saturation length larger than the system size corresponds to a purely erosional process. For many pratical situations --~e.g. formation of ripples and dunes~-- a first order relaxation of the type
\begin{equation}
L_{\rm sat} \partial_x q = q_{\rm sat} - q
\label{saturation}
\end{equation}
is a good approximation of the full dynamics.

In this paper, we review different aspects of subaqueous sediment transport: (i) the static threshold and (ii) the different regimes of transport. Our goal is to make clear the mechanisms responsible for these transport properties in order to gain understanding beyond the phenomenology. More precisely, the paper is constructed as follows. In the next section we compute the threshold Shields number $\Theta_{\rm th}$ from a miscroscopic and a continuum point of view. We discuss its dependency with longitudinal and transverse slopes, as well as cohesion. In section~\ref{erosionmomentumlimited} we describe the regimes in which transport is limited by the erosion process of the bed or momentum extraction by the grains from the fluid. In section~\ref{suspended} we discuss the case of suspended transport. In all these cases, we derive the expressions of the saturation flux $q_{\rm sat}$ and the saturation length $L_{\rm sat}$ and discuss their dependencies.

\section{Static threshold}
\label{Staticthreshold}

\subsection{Discrete description}
The bed load threshold is directly related to the fact that the grains are trapped in the potential wells created between neighbouring grains at the sand bed surface. To obtain the scaling law of the threshold shear stress, the simplest geometry to consider is a single spherical grain jammed between the two fixed grains below it. We first consider the situation in which the cohesive forces between the grains are negligible and the friction at the contacts is sufficient to prevent sliding.

The generic situation is a uniform shear stress $\rho_f u_*^2$, for which the velocity vertical profile is linear in the viscous regime:
\begin{equation}
u_x=\frac{u_*^2 z}{\nu} \, ,
\end{equation}
where $\nu$ is the kinematic viscosity of the fluid, and logarithmic in the turbulent regime:
\begin{equation}
u_x=\frac{u_*}{\kappa} \ln\left(1+\frac{z}{rd}\right),
\end{equation}
where $\kappa$ is the von K\'arm\'an constant and $r$ the ratio of the hydrodynamic roughness to the grain diameter ($z_0=rd$). The rescaled bed roughness $r$ is on the order of $r=1/10$. We introduce the effective velocity $U$ of the flow around the sand grains and define for convenience its normalised counterpart:
\begin{equation}
\mathcal{U} = \frac{U}{\sqrt{(\rho_s/\rho_f-1) gd}} \, .
\label{Uadim}
\end{equation}
For the sake of simplicity, we will take for $U$ the fluid velocity at the altitude $z=d/2$. Summing up viscous and turbulent contributions, a good approximation of the relation between the shear velocity $u_*$ and the typical velocity around the grains $U$ is given by
\begin{equation}
u_*^2=\frac{2 \nu }{d}  U+ \frac{\kappa^2}{\ln^2(1+1/2r)} U^2 \, .
\label{relationshearU}
\end{equation}
At the threshold, the horizontal force balance on a grain of the bed reads
\begin{equation}
\frac{\pi}{6}  \mu ( \rho_s - \rho_f ) g d^3= \frac{\pi}{8}  C_d  \rho_f U_{\rm th}^2 d^2,
\end{equation}
where $\mu$ is the avalanche slope for sand grains. $C_d$ is the drag coefficient, which is a function of the grain Reynolds number $\mathcal{R}U d/\nu$. In the viscous regime, the grain drag coefficient $C_d$ is inversely proportional to the grain Reynolds number ($C_d=s^2 \mathcal{R}^{-1}$); it is constant ($C_d=C_\infty$) in the turbulent regime. Note that the turbulent drag is not only due to the fluctuations induced by the grain itself, as in the case of a grain falling in a fluid at rest, but also to those present in the turbulent flow. This situation has never been studied properly so far. Note also that the drag force is {\it a priori} different in the vicinity of the ground and that the lift force could be non-negligible when the grain is trapped at the surface of the sand bed. In order to avoid introducing two many parameters, we will not consider these effects here. With a good accuracy, the drag law for natural grains can be put under the form:
\begin{equation}
C_d = \left(C_\infty^{1/2}+s \mathcal{R}^{-1/2} \right)^2 \quad{\rm with \quad} \mathcal{R}= \frac{U d}{\nu}\, .
\label{dragcoeff}
\end{equation}
Typical values found in sedimentation experiments performed with natural sand grains are $C_\infty \simeq 1$ and $s \simeq 5$.

\begin{figure}
\includegraphics{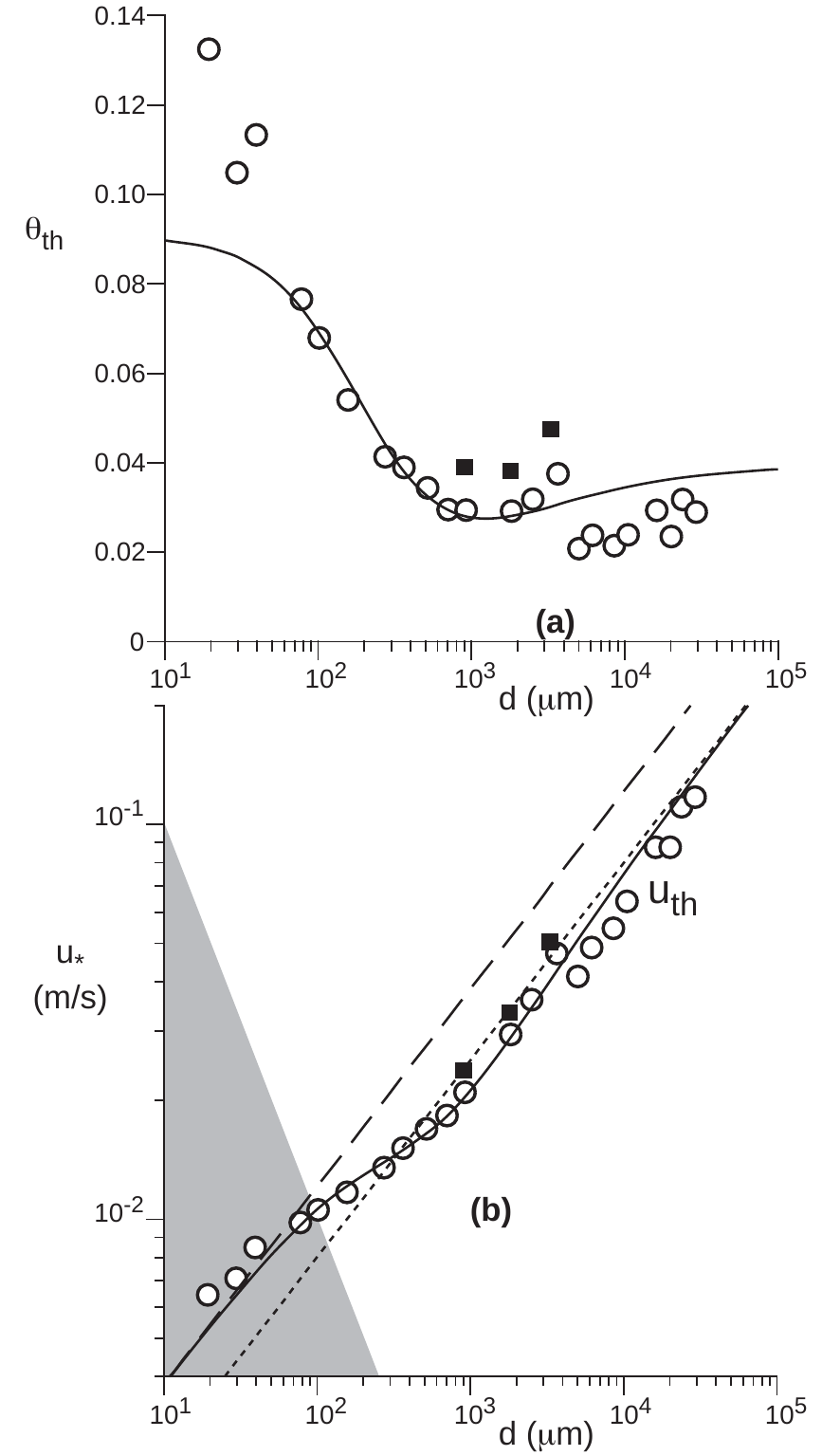}
\caption{(a) Threshold Shields number $\Theta$ as a function of the grain size $d$, for natural sand grains in water. Symbols: measurement by \cite{YK79} ($\circ$) and by \cite{LvB76} ($\blacksquare$). Solid line: Comparison with the model (equations~\ref{uth1}~and~\ref{uth2}).  (b) Threshold shear velocity $u_{\rm th}$ as a function of the grain size $d$, for natural sand grains. Symbols: measurement by \cite{YK79} ($\circ$) and by \cite{LvB76} ($\blacksquare$). Solid line: Comparison with the model. Dotted line: asymptotic behaviour in the turbulent regime. Dashed line: asymptotic behaviour in the viscous regime. Grey zone: zone in which the viscous sub-layer is larger than the bed roughness: for grains smaller than $d=100~\mu$m, the bed is hydraulically smooth for any shear stress $u_*$ above the threshold $u_{\rm th}$.
\label{Threshold}}
\end{figure}

At this stage, we introduce the viscous size $d_\nu$, defined as the diameter of grains whose
free fall Reynolds number $u_{\rm fall}d/\nu$ is unity:
\begin{equation}
d_\nu=(\rho_s/\rho_f-1)^{-1/3}~\nu^{2/3}~g^{-1/3}.
\end{equation}
It corresponds to a grain size at which viscous and gravity effects are of the same order of magnitude. From the three previous relations, we get the equation for $\mathcal{U}_{\rm th}$:
\begin{equation}
C_\infty^{1/2} \mathcal{U}_{\rm th} +s  \left(\frac{d_\nu }{d}\right)^{3/4}  \mathcal{U}_{\rm th}^{1/2} = \left(\frac{4\mu}{3}\right)^{1/2},
\end{equation}
which solves into
\begin{eqnarray}
\mathcal{U}_{\rm th} =\frac{1}{4 C_\infty}\left[\left( s^2\left(\frac{d_\nu }{d}\right)^{3/2} \right.\right. & + & 8 \left.\left(\frac{\mu C_\infty}{3}\right)^{1/2}\right)^{1/2} \nonumber \\
& - & \left. s  \left(\frac{d_\nu }{d}\right)^{3/4}\right]^2.
\label{uth1}
\end{eqnarray}
The corresponding expression of the static threshold Shields number is finally:
\begin{equation}
\Theta_{\rm th} =2 \left(\frac{d_\nu }{d}\right)^{3/2}  \mathcal{U}_{\rm th} +  \frac{\kappa^2}{\ln^2(1+1/2r)} \mathcal{U}_{\rm th}^2.
\label{uth2}
\end{equation}
In the viscous regime, the above relations simplify into:
\begin{equation}
\mathcal{U}_{\rm th}=\left(\frac{4 \mu}{3 s^2} \right) \left(\frac{d}{d_\nu}\right)^{3/2} \quad {\rm and} \quad \Theta_{\rm th} =\frac{8 \mu}{3 s^2} \, .
\end{equation}
In the turbulent regime,  we get:
\begin{equation}
\mathcal{U}_{\rm th}= \sqrt{\frac{4 \mu }{3 C_\infty}} \quad {\rm and} \quad \Theta_{\rm th} = \frac{4 \mu \kappa^2}{3 C_\infty \ln^2(1+1/2r)} \, .
\end{equation}
As can be seen in figure~\ref{Threshold}, the most interesting range of grain diameters, between $100~\mu$m and $1~$mm, is precisely the zone of transition between the two regimes. The crossover grain diameter $d_{co}$ scales as:
\begin{equation}
d_{co}=\frac{s^{4/3}}{(\mu C_\infty)^{1/3}}\,d_\nu,
\end{equation}
which is around $200~\mu$m.

The laminar-turbulent transition, indicated by the grey zone in figure~\ref{Threshold}b, is determined by the Reynolds number and thus depends on the grain diameter $d$. In most of the situations relevant in geophysics, $d$ lies in the cross-over between these two regimes. Figure~\ref{Threshold} shows the comparison between experimental data for natural sand grains and the full model. There is no adjustable parameter in the model as one can independently measure the avalanche slope $\mu=\tan 32^\circ$, the drag coefficients and the sand bed roughness $r=0.1$.
\begin{figure}
\includegraphics{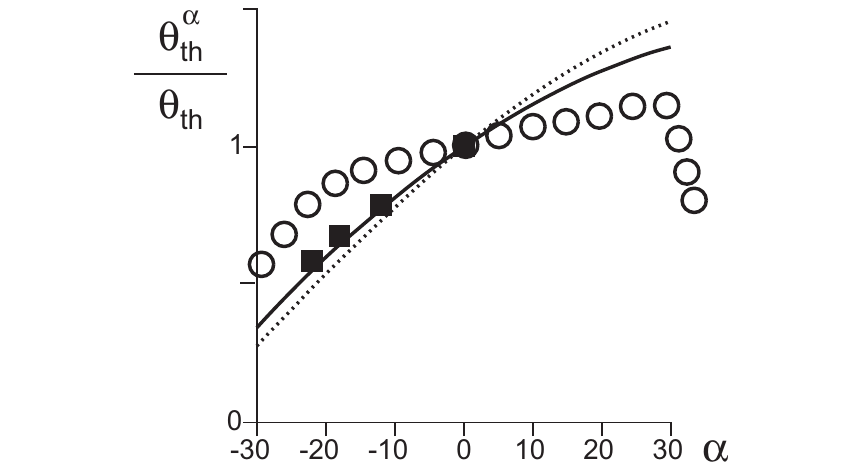}
\caption{Dependence of the threshold Shields number $\Theta_{\rm th}^\alpha$ on the bed inclination $\alpha$. Symbols: measurements by \cite{LvB76} for natural sand grains ($\blacksquare$) and by \cite{LGRD05} using glass beads ($\circ$) in a narrow channel. Solid line: expectation of the model for natural sand grains. Dotted line: approximation by $\cos \alpha+\sin \alpha/\mu$, with $\mu=\tan 32^\circ$.
\label{ThresholdSlope}}
\end{figure}
%

\subsection{Continuum description}
The transport threshold can alternatively be described using a continuum description of two-phase flows. We consider here the viscous limit. The turbulent case can be obtained in a straightforward way. In a steady state, the equilibrium equation reads:
\begin{eqnarray}
\rho_p \Phi \vec{g} +   \nabla \cdot \vec{\vec{\sigma}}^s +\Phi  \nabla \cdot \vec{\vec{\sigma}}^f +  \frac{3}{4}\,s^2\,\Phi \frac{\eta}{d^2}\vec{U}&=&\vec 0 \label{QMsolide2}\\
\rho_f (1-\Phi) \vec{g}+(1-\Phi)  \nabla \cdot \vec{\vec{\sigma}}^f - \frac{3}{4}\,s^2\,\Phi \frac{\eta}{d^2}\vec{U}&=&\vec 0  \label{QMfluid2}
\end{eqnarray}
where $\vec{\vec{\sigma}}^s$ and $ \vec{\vec{\sigma}}^f$ are respectively the sand borne and fluid borne stresses. Along the vertical direction, this equation reduces to the standard hydrostatic condition:
$$P^f=-\rho_f g z\quad{\rm et}\quad P^s=-\sigma^s_{zz}=-(\rho_p-\rho_f)\Phi g z$$
Along the horizontal direction, the force balance reads:
\begin{eqnarray}
\frac{d \sigma_{xz}^s}{dz} +\Phi  \frac{d \sigma_{xz}^f}{dz} + \frac{3}{4}\,s^2\, \Phi \frac{\eta}{d^2} U&=& 0 \label{QMsolide1}\\
(1-\Phi) \frac{d \sigma_{xz}^f}{dz} - \frac{3}{4}\,s^2\, \Phi \frac{\eta}{d^2} U&=&0 \label{QMfluid1}
\end{eqnarray}
with $ \sigma_{xz}^s=\mu \sigma_{zz}^s$ and $\sigma_{xz}^f=\eta U'(z)$ . Integrating the second equation, one gets the fluid velocity field:
$$u^f=\frac{\tau\,d}{\eta}\,\sqrt{\frac{4(1-\Phi)}{3\,s^2\, \Phi}}\,\exp\left(\sqrt{\frac{3\,s^2\, \Phi}{4\,(1-\Phi)}}\,\frac{z}{d}\right)$$
The fluid velocity relaxes exponentially to $0$ inside the sand bed, described as a porous medium. The first equation gives the sand borne shear stress:
$$\sigma_{xz}^s=\tau\,\left[1-\exp\left(\sqrt{\frac{3\,s^2\, \Phi}{4\,(1-\Phi)}}\,\frac{z}{d}\right)\right]$$
As expected, $\sigma_{xz}^s$ tends in depth to the total shear stress $\tau$. The Coulomb criterion $\sigma_{xz}^s/\sigma_{zz}^s=\mu$  is reached for the first time in $z=0$ and gives a threshold Shields number equal to:
$$\Theta_{\rm th}=\mu\,\sqrt{\frac{4 \Phi (1-\Phi)}{3\,s^2}}$$
For natural grains ($s \simeq 5$), the result is quantitatively very close to that given by the discrete approach ($\Theta_{\rm th} \simeq 0.07$), and to the measurements in the viscous regime.

\subsection{Effect of longitudinal slope}
Due to gravity, it is much easier to transport sediments along the lee side of a sand bump than on the stoss slope. This effect is directly encoded into the dependence of the saturated flux on the transport threshold, which depends on the slope. Let us note $\alpha$ the bed angle at the scale of the saturation length. The force balance at threshold now reads:
\begin{equation}
\frac{\pi}{6} ( \rho_s - \rho_f ) g d^3 \left(\sin \alpha+\mu \cos \alpha\right) = \frac{\pi}{8} C_d  \rho_f \left(U_{\rm th}^\alpha\,d\right)^2.
\end{equation}
We keep the notations $u_{\rm th}$ and $\Theta_{\rm th}$ for the threshold shear velocity and the threshold Shields number for an horizontal bed ($\alpha=0$) and denote by $u_{\rm th}^\alpha$ and $\Theta_{\rm th}^\alpha$ the thresholds in the case of an inclined bed. 

The dependence of the threshold on the slope has been measured experimentally by \cite{LvB76} for sand grains, in the turbulent regime. The model matches quantitatively the results, without any adjustable parameter (solid line on figure~\ref{ThresholdSlope}b). Neglecting the fact that $C_d$ is a function of the grain Reynolds number, the threshold shear stress can be written as: $\rho_f \left(u_{\rm th}^\alpha\right)^2=\rho_f u_{\rm th}^2 (\cos \alpha+\sin\alpha/\mu)$. This approximation still gives a reasonable fit of the data (dotted line in figure~\ref{ThresholdSlope}b).

The slope dependence of the threshold shear stress has also been measured in the case of aeolian transport, for rough sand grains (\cite{RIR96}) and is again perfectly fitted by the model. However, there exists an unexplained discrepancy in the case of spherical glass beads in a viscous liquid (Fig.~\ref{Threshold}b), as one would need a very low effective friction $\mu \sim \tan 70^\circ$ much lower than the avalanche slope ($\mu \sim \tan 24^\circ$) to fit the data (\cite{LGRD05}). With such a value, the gravity effect becomes completely negligible.

\subsection{Effect of transverse slope}

We consider a solid slide (a grain) on a surface $Z(x,y)$ inclined with a longitudinal slope $\partial_x Z$ and a transverse slope $\partial_y Z$. In the turbulent regime, for which the grain drag coefficient is roughly constant ($C_d \simeq 1$), the drag on a grain at the velocity $V \vec t$ reads:
\begin{equation}
\frac{\pi}{8}  C_d  \rho_f\,d^2\,|U \vec e_x-V \vec t|(U \vec e_x-V \vec t) ,
\end{equation}
and should balance the gravity and the solid friction:
\begin{equation}
-\frac{\pi}{6}\,\frac{( \rho_s - \rho_f ) g d^3}{\sqrt{1+(\partial_x Z)^2+(\partial_y Z)^2}}\,\left(\mu\,\vec t+\partial_x Z\,\vec e_x +\partial_y Z\,\vec e_y\right),
\end{equation}
At the threshold, the velocity $V$ vanishes so that:
\begin{equation}
\Theta_{\rm th}\,\vec e_x= \frac{4 \kappa^2}{3 C_d \ln^2(1+1/2r)}\,\,\frac{\left(\mu\,\vec t+\partial_x Z\,\vec e_x +\partial_y Z\,\vec e_y\right)}{\sqrt{1+(\partial_x Z)^2+(\partial_y Z)^2}},
\end{equation}
Defining the threshold Shields number for a flat bottom,
\begin{equation}
\Theta_{\rm th}^0 = \frac{4 \mu \kappa^2}{3 C_d \ln^2(1+1/2r)}\, ,
\end{equation}
we obtain:
\begin{equation}
\frac{\Theta_{\rm th}}{\Theta_{\rm th}^0}=\frac{\sqrt{1-\left(\partial_y Z\,/\mu\right)^2}+\partial_x Z\,/\mu}{\sqrt{1+(\partial_x Z)^2+(\partial_y Z)^2}}
\end{equation}
The threshold depends linearly on the longitudinal slope but quadratically on the transverse one.

\subsection{Effect of cohesion}

We now consider the case of cohesive granular media constituted of grains sufficiently small to lead to van der Waals interactions of the same order or larger than gravity. The adhesive force is related to the real surface of contact between grains. It depends on the loading history or, more precisely, on the maximum normal force exerted previously. When the cohesive grains have gently sedimented at the surface, this force is simply the weight of a single grain. On the contrary, when the grains have stayed under a very high pressure or are surrounded by clay, the whole surface contributes to the adhesion force. Introducing the surface tension $\gamma$, the maximum adhesion force is of the order of $\pi\,\gamma\,d$, which leads to a modified threshold Shields number:
\begin{equation}
\Theta^{\rm max}_{\rm th} (\gamma)=\Theta_{\rm th} (0)\left( 1+\frac{6\gamma}{\mu\,(\rho_s - \rho_f) g d^2}\right)
\end{equation}
The grain size at which the cohesive effects are of the same order as gravity is of the order of few millimeters in the case of capillary bridges and few tens of centimeters for clay.

\section{Bed load, from erosion to momentum limited regime}
\label{erosionmomentumlimited}

When a sand bed is submitted to a flow, only a small fraction of the grains at the surface are entrained and this erosion process takes some time to occur. Close to the threshold, there is a regime in which the concentration of mobile grains is not sufficient to induce a significant reduction of the flow strength in the transport layer. The grains transported in this dilute situation are isolated from each other. The saturation of the flux is then controlled by the erosion rate and the disorder of the sand bed. This idea is due to \cite{CME04}, who proposed a semi-phenomenological model that uses extensively experimental results.  As the concentration of transported grains becomes larger, the transport becomes limited by the available momentum: each time the flow entrains a grain from the bed and accelerates it, the grain exerts in return a stress on the fluid (\cite{B56}). The fluid in the transport layer is thus in equilibrium between the driving shear stress $\rho_f u_*^2$, the fluid-borne basal shear stress $\rho_f u_{\rm f}^2$ and this sand-borne shear stress. We propose here a model that includes these two regimes.

\subsection{Trajectory of a single grain}
\label{grainMotion}

\begin{figure}
\includegraphics{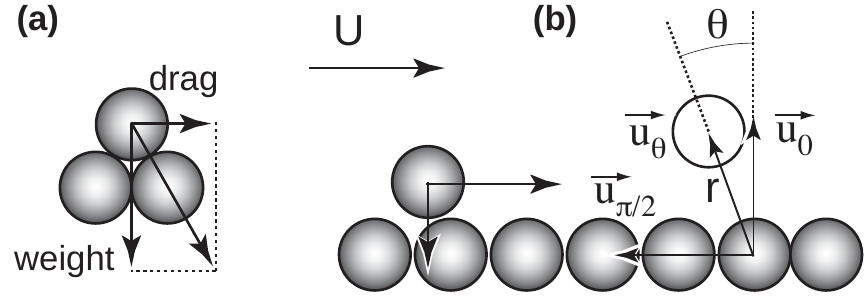}
\caption{Geometry of the trajectory calculation.}
\label{Trajectory}
\end{figure}

As for the computation of the threshold in the previous section, we shall start with the discrete description of the trajectory of a single grain over a regular bed (Fig.\ref{Trajectory}). To describe the elementary jump by one grain diameter, we write the equation of motion of one grain dragged by a flow of effective velocity $U$ and submitted to gravity. We assume that it looses completely its energy during collisions, due to the thin layer of fluid between the grains. Contrarily to \cite{B56} and \cite{CLDZ08}, we do not introduce any `effective friction': instead, we take into account the geometry of the trapping grains (Fig.~\ref{Trajectory}), which is the physical effect responsible for this friction.

Starting from the position of the grain $d \vec u_\theta$, we get its velocity $\vec V = d \dot \theta \vec u_{\theta+\pi/2}$ and its tangential acceleration $d \ddot \theta \vec u_{\theta+\pi/2}$. We assume that the reaction of the substrate is normal to it. Using again $d$ as a unit size and $[(\rho_s/\rho_f-1) gd]^{1/2}$ as a unit velocity, the dynamical equation reads:
\begin{eqnarray}
\ddot \theta & = & \frac{3}{4}  \left[C_\infty^{1/2}  \left(\dot \theta^2-2 \cos \theta\,\mathcal{U}\,\dot \theta +\mathcal{U} ^2\right)^{1/4} \right.
\nonumber\\
& + & s \left. \left(\frac{d_\nu }{d}\right)^{3/4}\right]^2
\left(\cos \theta\,\mathcal{U} -\dot \theta\right) + \sin \theta
\label{EqDyn}\\
& = & \left\{\frac{\sqrt{3}}{2}C_\infty^{1/2}  \left[\left(\dot \theta^2-2 \cos \theta\,\mathcal{U}\,\dot \theta +\mathcal{U}^2\right)^{1/4}-\mathcal{U}_{\rm th}^{1/2}\right] \right.
\nonumber\\
& + & \left. 3^{-1/4} \mathcal{U} _{\rm th}^{-1/2} \right\}^2  \left(\cos \theta\,\mathcal{U} -\dot \theta\right) + \sin \theta, \nonumber
\end{eqnarray}
where $\mu$ has been taken equal to $\tan(\pi/6)=1/\sqrt{3}$. This equation can be integrated numerically between $\theta=-\pi/6$ and $\theta=\pi/6$, starting from $\dot \theta=0$. Note that there is no adjustable parameter in this description.

In order to get the asymptotic behavior analytically, we perform an expansion with respect to $\theta$ around $-\pi/6$ and with respect to $\mathcal{U}$ around $\mathcal{U}_{\rm th}$. We get a linear equation of the form:
\begin{eqnarray}
\ddot \theta + \left(\frac{1}{\sqrt{3}\,\mathcal{U}_{\rm th}}+\frac{3^{5/4}}{8} C_\infty^{1/2}\right)  \dot \theta
& - & \frac{2}{\sqrt{3}} \left(\theta+\frac{\pi}{6}\right)
\nonumber \\
& = & \frac{2\pi \left(\mathcal{U} - \mathcal{U}_{\rm th} \right)}{3^{3/2} a \mathcal{U}_{\rm th}} \,,
\end{eqnarray}
with
\begin{equation}
a=\frac{8 \pi }{3^{3/2} \left(2+3^{3/4} C_\infty^{1/2} \mathcal{U}_{\rm th} \right)} \,
\nonumber
\end{equation}
The solution of this linear equation is:
\begin{equation}
\theta=-\frac{\pi}{6} + \frac{\pi \left(\mathcal{U}_{\rm th}-\mathcal{U}\right)}{3 a \mathcal{U}_{\rm th}} \left[1+\frac{r_- e^{r_+t}-r_+e^{r_- t}}{r_+-r_-}\right],
\end{equation}
where $r_-$ and $r_+$ are the solutions of the equation:
$$r^2 +  \left(\frac{1}{\sqrt{3} \mathcal{U}_{\rm th} }+\frac{3^{5/4}}{8} C_\infty^{1/2}\right)  r - \frac{2}{\sqrt{3}}=0. $$
The time ${\mathcal T}$ needed for a grain to move by one diameter thus scales around the threshold as:
\begin{equation}
{\mathcal T} \sim \frac{1}{r_+}\,\ln \left[ \left(1-\frac{r_+}{r_-}\right) \left(1+\frac{a \mathcal{U}_{\rm th}}{\mathcal{U}-\mathcal{U}_{\rm th}} \right) \right]\,\sqrt{\frac{\rho_f d}{(\rho_s-\rho_f) g}} \, .
\label{ExpT}
\end{equation}
This erosion time diverges at the threshold as:
\begin{equation}
{\mathcal T} \propto\,-\ln \left[\Theta-\Theta_{\rm th} \right].
\end{equation}
On the other hand, ${\mathcal T}$ should resume to $d/U$ far from the threshold. Integrating numerically equation~(\ref{EqDyn}), we have found convenient approximations that follow the correct asymptotic behaviors just above the threshold and far from it. In the turbulent regime, it reads:
\begin{equation}
{\mathcal T} \propto\,\ln \left[\frac{\sqrt{\Theta}+\sqrt{\Theta_{\rm th}}}{\sqrt{\Theta}-\sqrt{\Theta_{\rm th}}} \right]\,\sqrt{\frac{\rho_f d}{(\rho_s-\rho_f) g}} \, ,
\label{ApproxTturb}
\end{equation}
where the prefactor is around $1.5$. In the viscous regime, it reads
\begin{equation}
{\mathcal T} \propto\,\ln \left[\frac{\Theta+\Theta_{\rm th}}{\Theta-\Theta_{\rm th}} \right]\, \left(\frac{d_\nu }{d}\right)^{3/2}  \,\sqrt{\frac{\rho_f d}{(\rho_s-\rho_f) g}} \, ,
\label{ApproxTvisc}
\end{equation}
with a prefactor around $17$.
\begin{figure}
\includegraphics{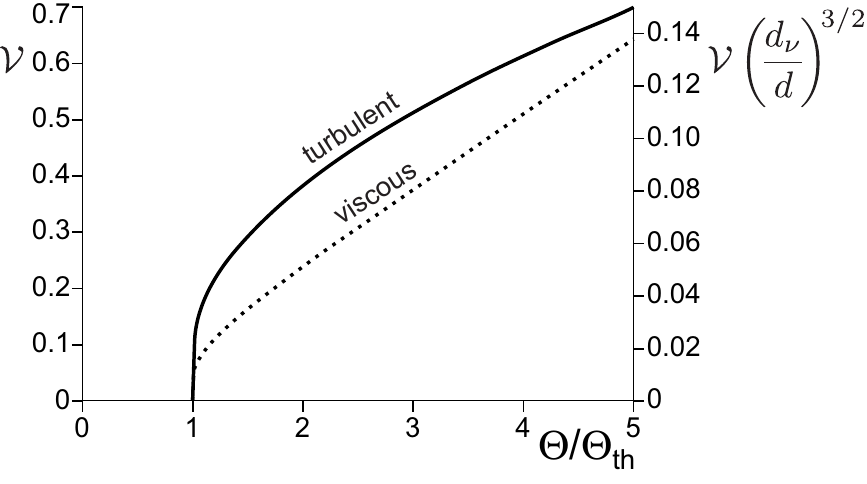}
\caption{Left: mean grain velocity ${\mathcal V}=\left((\rho_s/\rho_f-1) gd\right)^{-1/2}\,\frac{d}{\mathcal T}$ as a function of the rescaled Shields number $\Theta/\Theta_{\rm th}$ in the turbulent regime  (large $d$, solid line). Right: same, but  in the viscous regime (small $d$, dotted line), with an further factor $\left(\frac{d_\nu }{d}\right)^{3/2}$ .
\label{ErosionTime}}
\end{figure}
%

\subsection{Saturated flux}

We define $\mathcal{N}(\Theta)$ the fraction of grains at the surface susceptible to be entrained at a Shields number $\Theta$. In the absence of flow, for $\Theta=0$, $\mathcal{N}$ is null. It increases very quickly around the mean threshold Shields number $\Theta_{\rm th}$ and reaches $1$ at a second threshold $\Theta_M$.  $\mathcal{N}(\Theta)$ reflects and encodes the distribution of potential wells at the sand bed surface.  A simple choice is:
\begin{eqnarray}
\mathcal{N}&=&0\quad {\rm if} \quad \Theta<\Theta_m, \nonumber\\
\mathcal{N}&=&\frac{\Theta-\Theta_m}{\Theta_M-\Theta_m}\quad {\rm if} \quad \Theta_m<\Theta<\Theta_M, \\
\mathcal{N}&=&1\quad {\rm if} \quad \Theta>\Theta_M.\nonumber
\end{eqnarray}

As suggested by the experiments of \cite{CME04}, a granular bed prepared by sedimentation is initially very disordered and consequently $\Theta_M$ is well above $\Theta_{\rm th}$ and large. It then takes a very long time for the surface to re-arrange, leading to a drift of the distribution $\mathcal{N}$ toward larger and larger threshold shear stresses. This long transient is probably not relevant to geophysical situations in which the time and length scales of the systems are always sufficiently large to ensure that an equilibrium state for the geometrical arrangement of surface grains has been reached.

When one grain is entrained, we hypothesise that there can be no further erosion in a surrounding area $d^2$ until the grain has moved by a distance comparable to its own diameter $d$. The erosion time ${\mathcal T}$ previously computed leads to a mean velocity that goes to $0$ at the threshold like $1/\ln \left[\Theta-\Theta_{\rm th} \right]$. This corresponds to a bifurcation much sharper than the standard saddle node bifurcation in $\sqrt{\Theta-\Theta_{\rm th}}$. The experimental observation (\cite{AF77,LvB76,CME04}) that the grain velocity --~i.e. the mean velocity conditioned by the fact that the grain moves significantly at the time resolution of the instrument~-- is non zero just above the threshold is thus consistent with this expression. Far from threshold, the grain velocity is just proportional to the flow velocity (\cite{AF77,LvB76,CME04}). The length of a trajectory is independent of the trap in which the grain was initially at rest and fluctuates around a well defined average value ${\mathcal L}$. The mean area explored by the grain is ${\mathcal L} d$ and contains, by definition of ${\mathcal L}$, a mean number of potential wells sufficiently deep to trap the grain equal to $1$. We then obtain the relation: ${\mathcal L} \left(1-\mathcal{N}(\Theta)\right)=d$.

The particles are entrained by the flow at a shear velocity $u_{\rm f}$ reduced by the presence of other particles. The relevant Shields number $\Theta$ should thus be based on $u_{\rm f}$ rather than $u_*$, to take the negative feedback of the transport on the flow into account.  We consider the regime in which the grains are transported at the surface of the bed i.e. do not form a surface sheet flow (\cite{OAG09a}). They leave the bed with a velocity $v_\uparrow$ and collide back the sand bed with a velocity $v_\downarrow$. The sand-borne shear stress is proportional to the sand flux and to the difference ($v_\downarrow-v_\uparrow$). The equilibrium of the fluid in the transport layer leads to the relation valid in the saturated state ($q=q_{\rm sat}$) but also during the transient of saturation
\begin{equation}
\rho_f u_*^2=\rho_f u_{\rm f}^2+ \rho_s \phi \frac{(v_\downarrow-v_\uparrow)}{{\mathcal L}} \, q.
\label{qq0}
\end{equation}
The grain trajectory is encoded into a single quantity, $(v_\downarrow-v_\uparrow)/{\mathcal L}$, which is a function of the reduced shear velocity $u_{\rm f}$ and \emph{not} of $u_*$.
\begin{figure}
\includegraphics{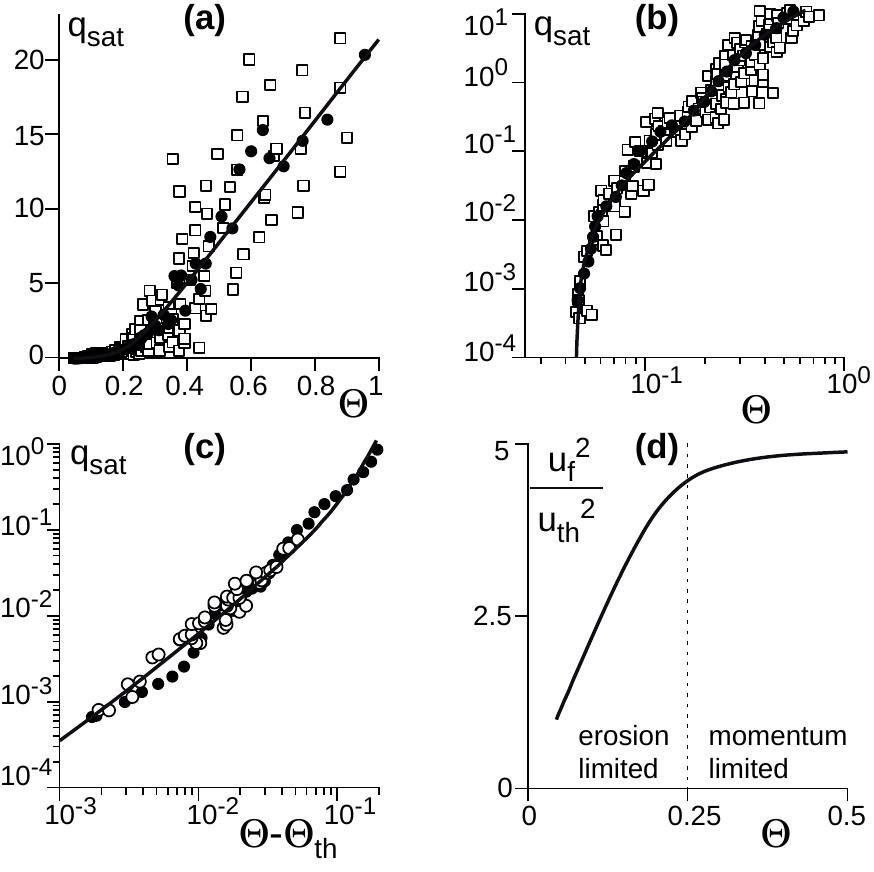}
\caption{(a-c) Saturated flux $q_{\rm sat}$ rescaled by $\sqrt{\left(\frac{\rho_s}{\rho_f}-1\right) g d^3}$ as a function of the Shields number $\Theta$: raw data ($\square$) gathered by \cite{J98}; same after local averaging ($\bullet$); raw data ($\circ$) obtained by \cite{LvB76}. The solid line is the best fit by the model described here (equations \ref{qqq1} and \ref{qqq3}), which gives: $\Theta_{\rm th}=0.045$ and $\Theta_{\rm M}=5\,\Theta_{\rm th}$. (a) Lin-lin plot; (b) Log-log plot; (c) Log-log plot as a function of  $\Theta-\Theta_{\rm th}$. (d) Velocity $u_f$ in the transport layer as a function of $\Theta$ determined from the model (equations \ref{qqq1} and \ref{qqq3}).
\label{FluxSature}}
\end{figure}

The saturated flux $q_{\rm sat}$ is simply the erosion rate times the hop length ${\mathcal L}=d/\left(1-\mathcal{N}(\Theta)\right)$. It can be expressed as an integral over all possible local configurations:
\begin{equation}
q_{\rm sat}= \frac{\pi d^2}{6 \phi \left(1-\mathcal{N}(\Theta)\right)} \int_0^{\Theta}  \frac{\mathcal{N}'\left(\Theta_{\rm th}\right)}{{\mathcal T}(\Theta_{\rm th},\Theta)} d \Theta_{\rm th},
\end{equation}
where $\phi$ is the bed volume fraction. A good approximation of this integral is given by:
\begin{equation}
q_{\rm sat}\propto \frac{\left( u_{\rm f}^2-u_{\rm th}^2 \right)}{\phi \, \left(u_{\rm M}^2-u_{\rm f}^2 \right)\,\left[c_m+\ln \left(\frac{u_{\rm f}^2}{u_{\rm f}^2-u_{\rm th}^2} \right)\right] }\,\sqrt{\left(\frac{\rho_s}{\rho_f}-1\right) g d^3}.
\label{qqq1}
\end{equation}
where $c_m\simeq1.41$ in the turbulent regime and $c_m\simeq0.605$ in the viscous regime. In particular one can recover from this expression the expected asymptotic behaviours. By construction, it vanishes at the shear velocity $u_{\rm th}$ and diverges at $u_M$. For the sake of simplicity, we limit the discussion to the turbulent case. The transposition to the viscous case is straightforward. Assuming that the velocity increment $v_\downarrow-v_\uparrow$ scales on the grain mean velocity, the ratio $(v_\downarrow-v_\uparrow)/{\mathcal L}$ is proportional to the hop time ${\mathcal T}$, which is also the erosion time. Equation~(\ref{qq0}) can be solved to get the flux:
\begin{eqnarray}
q_{\rm sat} & \propto & \frac{\rho_f}{\rho_s}\, \left(u_*^2-u_{\rm f}^2\right)\,{\mathcal T}(u_{\rm f}) \nonumber\\
& \propto & \frac{\rho_f}{\rho_s}\,\sqrt{\frac{\rho_f d}{(\rho_s-\rho_f) g}} \, \ln \left[\frac{u_{\rm f}+u_{\rm th}}{u_{\rm f}-u_{\rm th}} \right] \, \left(u_*^2-u_{\rm f}^2\right).
\label{qqq2}
\end{eqnarray}
Eliminating the flux between the equations (\ref{qqq1}) and (\ref{qqq2}), one obtains a relation between $u_{\rm f}$ and $u_*$.
\begin{eqnarray}
\left(u_*^2 \right.&-& \left. u_{\rm th}^2\right) \propto \left(u_{\rm f}^2-u_{\rm th}^2 \right) \nonumber\\
& \times &
\left[1+
\frac{\frac{\rho_s}{\rho_f}\,\left(\frac{\rho_s}{\rho_f}-1\right) g d}{\phi \, \left(u_{\rm M}^2-u_{\rm f}^2 \right)\,\left[c_m+\ln \left(\frac{u_{\rm f}^2}{u_{\rm f}^2-u_{\rm th}^2} \right)\right]\,\ln \left[\frac{u_{\rm f}+u_{\rm th}}{u_{\rm f}-u_{\rm th}} \right] }\right].
\label{qqq3}
\end{eqnarray}

Just above the threshold, the flow in the transport layer is undisturbed: $u_{\rm f} \sim u_*$. This corresponds to the erosion limited regime and the flux can be approximated by:
\begin{equation}
q_{\rm sat}\propto \frac{1}{\phi \,\left[c_m+\ln \left(\frac{u_*^2+u_{\rm th}^2}{u_*^2-u_{\rm th}^2} \right)\right]}
\left( \frac{u_*^2-u_{\rm th}^2}{u_{\rm M}^2-u_*^2} \right)
\sqrt{\left(\frac{\rho_s}{\rho_f}-1\right) g d^3}
\end{equation}

Far above the threshold, the velocity profile inside the transport layer  becomes independent of $u_*$.  The shear velocity $u_{\rm f}$ tends to $u_{\rm M}$, value for which all the grains at the surface can be mobilised. The flux then scales as:
\begin{equation}
q_{\rm sat} \propto \frac{\rho_f}{\rho_s}\,\sqrt{\frac{\rho_f d}{(\rho_s-\rho_f) g}} \, \left(u_*^2-u_{\rm M}^2\right)
\end{equation}
Thus, in the momentum limited regime, $q$ scales at large velocities as $u_*^2$ and not $u_*^3$ as usually obtained. The same result is valid for aeolian transport, for which the negative feedback of particles on the flow has been directly evidenced experimentally. Although many authors have followed Bagnolds, the scaling law in $u_*^2$ is, in that case, the best description of existing data \cite{A04}. In figure~\ref{FluxSature}, we have re-plotted the measurements of the saturated flux collected by \cite{J98} for the subaqueous case. It exhibits a large dispersion due to several factors. First, data coming from systems with different grain size distributions have been plotted together without any distinction of symbols. Second, the reproducibility of experiments is made difficult by the problem of granular bed preparation: we emphasise again that sedimentation leads to an out of equilibrium situation that can last for days. Third, it is difficult to estimate the basal shear stress in flume experiments due to lateral boundaries. We have also plotted the measurements performed by \cite{LvB76}, which are much less scatterred. Given these reservations, the fit by the model derived here provides a good description of existing data. In particular, the asymptotic behaviors close to the threshold and for large shear velocities are well captured.

\subsection{Saturation length}
The sand flux is the product of the grain velocity by the concentration of mobilised grains. Two important mechanisms can thus control the saturation length: the length needed by the grain to reach its asymptotic velocity --~the so-called drag length~-- and the length needed for erosion to take place i.e. the trajectory length ${\mathcal L}$. The modelling of the drag length is a difficult problem as the trajectory takes place in a turbulent flow whose fluctuations are \emph{not} due to the motion of the grain itself. The problem is thus very different from that of a sphere moving in a fluid at rest, a problem for which the drag law is calibrated. To the best of our knowledge, the motion of a sphere whose diameter lies in the inertial range of the turbulent flow is still an open problem. We are thus left with the standard drag force formula $\frac{\pi}{8} C_d \rho_f U^2 d^2$, with a drag coefficient of order one. Then, solving the equation of motion, one obtains a drag length $L_{\rm drag}$ around $2\,(\rho_s/\rho_f)d$.

In summary, we consider here that $L_{\rm sat}$ can be either limited by erosion, in which case it is expected to scale as:
\begin{equation}
L_{\rm sat} =\frac{d}{1-\mathcal{N}(\Theta)} \, .
\label{LsatAqueux}
\end{equation}
At small $\Theta$, $L_{\rm sat}$ is simply one grain diameter $d$; at $\Theta_M$, $\mathcal{N}$ tends to $1$ so that $L_{\rm sat}$ would diverge. The divergence of the trajectory length has been directly evidenced experimentally in the viscous case, by \cite{CLDZ08}. However, the saturation length becomes limited by the grain inertia below $\Theta_M$, in which case we get:
\begin{equation}
L_{\rm sat} \simeq 2 \frac{\rho_s}{\rho_f} d
\end{equation}
One can see that these two predictions are difficult to test and discriminate. The erosion length gently increases with the shear velocity while the drag length is independent of it; the drag length increases with the density ratio $\rho_s/\rho_f$ but, experimentally, this parameter cannot be easily varied by a large factor.

\section{Suspended load}
\label{suspended}
We consider a flow of thickness $H$ above an erodible bed. We introduce $\phi$, the mass  concentration of particles in suspension. We limit the discussion to dilute turbulent suspensions, in which there is no significant feedback of the particles on transport. Assuming that the suspended sediment transport is linearly driven by the concentration gradient through a particle turbulent diffusion coefficient $D$, the equation governing the evolution of $\phi$ reads
\begin{equation}
\dt \phi + u_x \dx \phi = \dx \left(D \dx \phi \right) + \dz \left(D \dz \phi + \phi V_{\rm fall} \right),
\label{eq:concentration}
\end{equation}
where $V_{\rm fall}$ is the settling velocity, assumed uniform and equal to that in the dilute limit. Note that we neglect here the vertical component of the advection. The sediment flux $q$ introduced above is here defined as
\begin{equation}
q = \frac{1}{\rho_s} \int \phi u_x dz.
\label{eq:defq}
\end{equation}
For simplicity, we consider a flow with velocity $U$ assumed uniform in the vertical $z$-direction. We assume the friction velocity $u_*$ is assumed to be proportional to the flow velocity, i.e. $U =\lambda u_*$. The diffusion flux is assumed to be proportional to the gradient of the particle concentration, with diffusivity $D = \mathcal{K} u_* H$, where $\mathcal{K}$ is a constant.

In the saturated state, steady and homogeneous in the longitudinal direction, the vertical diffusive flux of particles is balanced by the settling flux. With the boundary condition of zero flux at the free surface, equation~(\ref{eq:concentration}) governing the evolution of the concentration then reduces to
\begin{equation}
D \frac{\partial \phi}{\partial z} + \phi\,V_{\rm fall} = 0,
\label{steadyandhomogeneous}
\end{equation}
so that the equilibrium particle density, denoted $\Phi$, decreases exponentially with height:
\begin{equation}
\Phi=\Phi_0\,\exp\left(-\alpha \frac{z}{H}\right) \quad{\rm with}\quad \alpha\equiv\frac{V_{\rm fall}}{\mathcal{K} u_*}.
\label{eq:Phi}
\end{equation}
In a thin layer close to the bed, of thickness of a few particle diameters, the settling velocity vanishes and the particle concentration increases to that of the fixed bed, $\phi_M \approx 0.6$. Just above this layer, which may be considered as the bedload layer, say at $z=0$, continuity of the erosion flux gives the boundary condition
\begin{equation}
\varphi(\tau) = - D \frac{\partial \phi}{\partial z} 
\qquad{\rm at} \quad z=0,
\label{eq:BC@z=0}
\end{equation}
which determines the expression of the concentration $\Phi_0$:
\begin{equation}
\Phi_0 = \frac{\varphi(\tau)}{V_{\rm fall}}.
\end{equation}
Finally, the saturated particle flux per unit width, normalized by the water flux $U H$, is given by
\begin{equation}
\frac{q_{\rm sat}}{U H} = \frac{1}{U H} \, \int_0^H \Phi U dz = 
\frac{1 - {\rm e}^{-\alpha}}{\alpha} \, \Phi_0
\end{equation}
with the assumption that particles have negligible inertia and follow the stream. For small $\alpha$, this dimensionless flux tends to $\Phi_0$, as expected.

The saturation length is controlled in this regime by the length needed by a particle at a typical height $\simeq H$ above the bed to settle down:
\begin{equation}
L_{\rm sat} \propto \frac{u_*}{V_{\rm fall}} \, H.
\end{equation}
%


\end{document}